\documentclass[twocolumn,aps,showpacs,pra]{revtex4}
\usepackage{epsfig}
\usepackage[english]{babel}
\usepackage{latexsym}
\usepackage{graphics}
\usepackage{subfigure}
\usepackage{epsfig}
\usepackage{graphicx}
\usepackage{dcolumn}
\usepackage{amsmath}
\begin{document}
\title{High-order Harmonic Generation in a Time-integrated Quantum Transition Picture}
\author{Y. J. Chen$^{1,2}$,  J. Liu$^{1,3*}$, and Bambi Hu$^{2,4}$}

\date{\today}

\begin{abstract}
We develop a  numerical scheme to investigate the high-order
harmonic generation (HHG) in intense laser-matter interactions.
Tracing the time evolution of every electronic laser-field-free
state, we  "observe" the HHG in a time-integrated quantum transition
picture. Our full-quantum simulations reveal that continuum
electrons with a broad energy distribution contribute equally to one
harmonic  and the excited state also plays an important role in the
molecular HHG. These results imply a laser-intensity-dependent
picture of intramolecular interference in the  HHG.

\end{abstract}
\affiliation{1.Institute of Applied Physics and Computational
Mathematics, Beijing 100088, China\\2. Department of Physics, Centre
for Nonlinear Studies, and The Beijing-Hong Kong-Singapore Joint
Center for Nonlinear and Complex Systems
(Hong Kong), Hong Kong Baptist University, Kowloon Tong, Hong Kong, China\\
3.Center for Applied Physics and Technology, Peking University,
Beijing 100084, China\\4. Department of Physics, University of
Houston, Houston, Texas 77204-5005.} 
\maketitle

Owing to the important applications in producing burst of high
energy photon, high-order harmonic generation (HHG) is a main focus
of intense laser-matter physics\cite{M. Hentschel}. In recent years,
it is also shown that the HHG can be used to image molecular
orbital\cite{ilzn04}. The HHG can be well understood by a
semiclassical recollision model\cite{Corkum}: (i) ionization of the
active electron by tunnelling; (ii) acceleration of the electron in
the laser field; (iii) recombination of the electron into the bound
state to emit a high energy photon. Another widely used approach to
describe the HHG is the strong field approximation
(SFA)\cite{Lewenstein}, which can be regarded as the
quantum-mechanism version of the semiclassical recollision model.

Numerical investigation on the HHG of H$_2^+$ demonstrates an
interesting phenomenon: a minimum appears in the HHG spectrum that
relates to the molecular orientation. Using a point-emitter model,
the minimum is identified as arising due to intramolecular
two-center interference and the position of the minimum is concluded
not to  rely on the laser intensity\cite{lein1}. This minimum has
generated great theoretical and experimental
interests\cite{Bandrauk2,Tsuneto,vozzi,Anh-Thu Le,cyj2,Kanai,Xibin
Zhou,Ciappina,Baker}. Two experimental groups reported the
observations of the minima in the HHG spectra of
CO$_2$\cite{Tsuneto,vozzi}. They attributed the minima to two-center
interference. However, the positions of the minima are different in
their measurements (the 27th vs the 33rd orders). Other than
interference, the theoretical calculations proposed that the ground
state depletion is also a possible mechanism for the different
modulations observed\cite{Anh-Thu Le}.  Zhou et al showed that
two-center interference is responsible for the  minima in the HHG
spectra of CO$_2$ measured in their experiments\cite{Xibin Zhou}.
Nevertheless, they found that the interference pattern is subjected
to the Coulomb effects. Very recently, the experimental observations
of the interference minima in the HHG spectra of H$_2$ were
reported\cite{Baker}. But the minima are found to be related
to %dependent on
the laser intensity. This relation is  attributed to the motion of
the parent nuclei there. All these strongly call for a revisit of
the HHG, especially,  a quantitative description of intramolecular
interference in the HHG.

The core of the HHG is the recombination process and the
recombination is an intermediate process that can not be observed
experimentally. In addition, the semiclassical models that  omit the
Coulomb effects can not describe the HHG quantitatively. Thus, in
this paper, we develop a numerical scheme to investigate the
process. We directly solve the Born-Oppenheimer time-dependent
Schr\"odinger equation (TDSE), and we project the wavefunction on
the eigen-states of the field-free Hamiltonian and trace its
temporal evolution. With
the method, we achieve a time-integrated %"static"
quantum transition picture of the HHG. It gives us direct
information about the elaborate role of every electronic state in
the HHG and allows us to identify the whole energy transfer among
harmonics, the laser field and the electrons in the HHG. Our
simulations reveal that the major contributions to one harmonic come
from continuum electrons with a  broad energy distribution and the
first excited state  also has a significant contribution to the
molecular HHG. As a result, the position of the interference minimum
is affected by the laser intensity in the frozen nuclei case.

Below, our discussion is first made for the 1D  case, we then
validate the main results to the 2D and 3D cases. The Hamiltonian of
model molecules H$_2^+$ or hydrogen-like atoms, all with the
ionization potential $I_p=1.11$ a.u. studied here, is
$H(t)=\mathbf{p}^2/2+V(\mathbf{r})-\mathbf{r}\cdot \mathbf{E}(t)$
%(atomic units of $\hbar=e=m_e=1$ are adopted).
(in a.u. $\hbar=e=m_e=1$). Here, $V(\mathbf{r})$ is the Coulomb
potential. In the 1D and 2D cases, we use the soft-Coulomb potential
$V(x)=\frac{-Z}{\sqrt{1.44+({x+R/2})^{2}}}+\frac{-Z}{\sqrt{1.44+({x-R/2})^{2}}}$
and $ V(x,y)=\frac{-Z}{\sqrt{0.5+({x+R/2})^{2}+y^2}}+
\frac{-Z}{\sqrt{0.5+({x-R/2})^{2}+y^2}}$, respectively, where  $Z$
is the effective charge, $R$ is the internuclear separation. For
H$_2^+$, $R=2$ a.u..  $R=0$ a.u. corresponds to the hydrogen-like
atom. In the case of the 3D atom, it is $V(\mathbf{r})=-Z/r$.
%where $Z$ is the effective charge.
$\mathbf{E}(t)=\vec{\mathbf{e}}\mathcal{E}\sin{\omega_{0}}t$
is the external electric  field         %
with the amplitude $\mathcal{E}$ and the frequency $\omega_{0}$.
$\vec{\mathbf{e}}$ is the unit vector along the laser polarization.
We use $\theta$ to denote the angle between the molecular axis and
the laser polarization. Our calculation will be performed for $780$
nm trapezoidally shaped laser pulses with a total duration of $10$
optical cycles and linear ramps of three optical cycles. The details
of solving the 1D and 2D TDSE can be found in Ref.\cite{cyj}. The 3D
TDSE is solved following the program described in Ref.\cite{milo}.
The coherent part of the HHG spectrum is obtained by\cite{lein1}
\begin{eqnarray}
F(\omega)=\int\langle\psi(t)|\vec{\mathbf{e}}\cdot\nabla
V|\psi(t)\rangle e^{i\omega t}dt,
\end{eqnarray}
where $|\psi(t)\rangle$ is the time-dependent wavefunction of
$H(t)$,  and $\omega$ is the emitted proton frequency.

%%%%%%%%%%%%%%%%%%%%%%%%%%%%%%%%%%%%%%%%%%%%%%%%%%%%%%%%%%%%%%%%%%%%%%%%%%%%%%%%%%%%%%%%%
\begin{figure}[t]
\begin{center}
\rotatebox{0}{\resizebox *{8.5cm}{8.0cm} {\includegraphics
{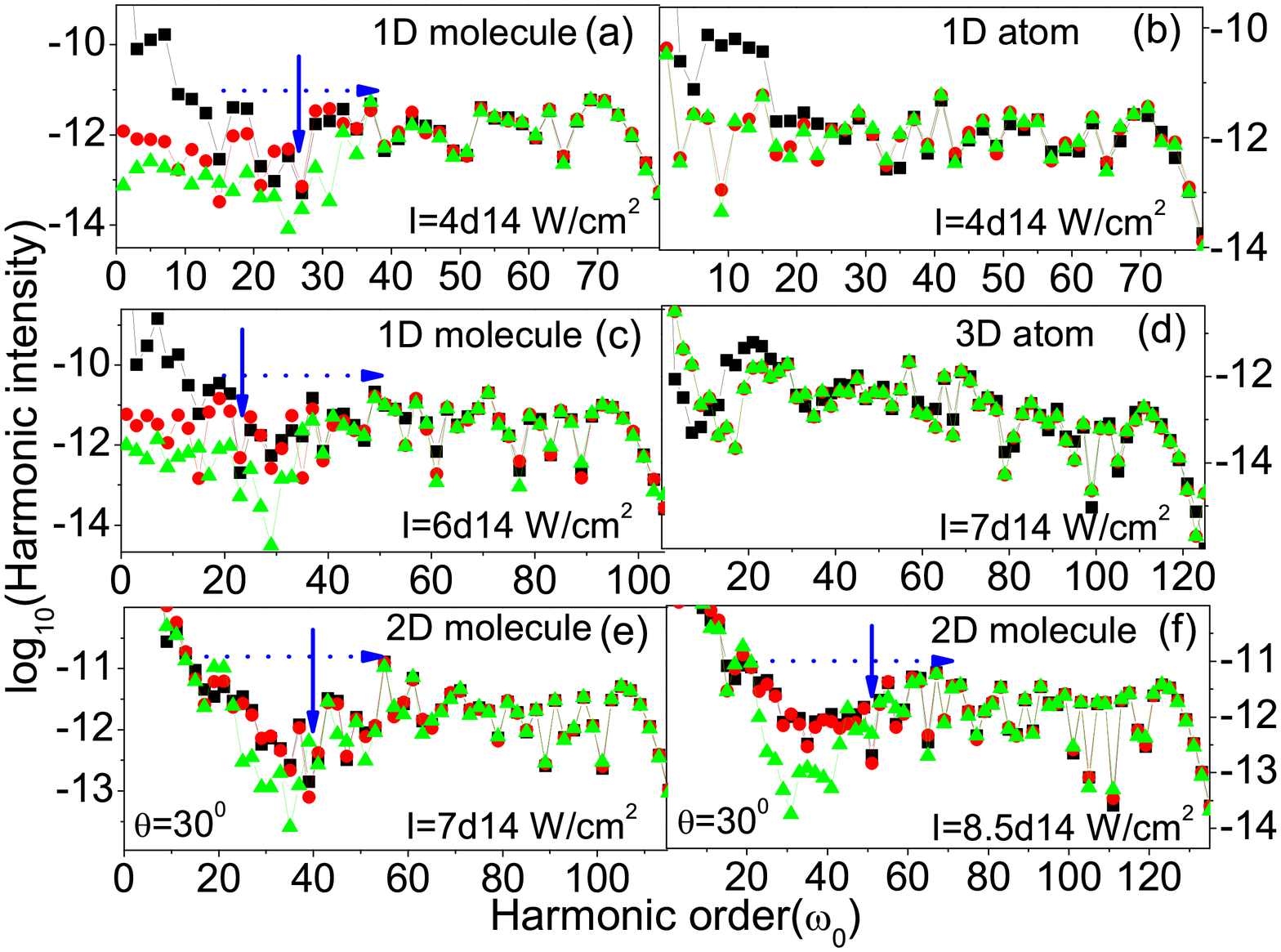}}}
\end{center}
\caption{Harmonic spectra of H$_2^+$ molecules [(a),(c),(e),(f)]
and hydrogen-like
atoms [(b),(d)] %at the field intensity $I$$=$$4\times10^{14}\mathrm{W/cm^2}$.
calculated using different methods. The black curves denote the
results obtained using Eq. 1, the red and green curves denote those
obtained using Eq. 2 with $n'=0,1$ and $n'=0$, respectively. For the
2D and 3D cases in (d)-(f), also see the context for the details. }
\label{Fig. 1}
\end{figure}
%%%%%%%%%%%%%%%%%%%%%%%%%%%%%%%%%%%%%%%%%%%%%%%%%%%%%%%%%%%%%%%%%%%%%%%%%%%%%%%%%%%%%%%%%%

To achieve a full-quantum description of the recombination in the 1D
case, we project the wavefunction $|\psi(t)\rangle$ on the
eigen-states of the field-free Hamiltonian
$H_{0}=\mathbf{p}^2/2+V(x)$, that is $
|\psi(t)\rangle=\sum_na_{n}(t)|n\rangle+\int
d\mathbf{p}c_{\mathbf{p}}(t)|\mathbf{p}\rangle. $ Here,
$\psi_{c}(\mathbf{r},t)=\int
d\mathbf{p}c_{\mathbf{p}}(t)|\mathbf{p}\rangle$ represents the
continuum electronic wave packet. $|n\rangle$ is the bound state
%eigen-wavefunction of the bound states
of $H_0$ with negative eigen-energy $E_n$; $|\mathbf{p}\rangle$
represents the continuum state of $H_0$ that has positive
eigen-energy denoted by $E_\mathbf{p}$. The eigen-states and
eigen-values are obtained by numerically diagonalizing the
Hamiltonian $H_{0}$.

The recombination  in the HHG can be  simulated by the evaluation of
the dipole movement between the "concerned" bound states and the
continuum electronic wave packet $\psi_{c}(\mathbf{r},t)$, i.e., $
D(t)=\sum_{n'}a^*_{n'}(t)\langle n'|\nabla
V|\psi_{c}(\mathbf{r},t)\rangle. $ Here, $n'$ indicates the
concerned bound state that the continuum electrons can transit back
to. Because the population of  other higher bound states is found to
be negligible in our simulations. In this paper, we focus on the
dipole movement in the cases of $n'=0$ and $n'=0,1$, i.e., the
dipole movement relating to the transition back to the ground state
$|0\rangle$ and to both the ground state $|0\rangle$ and the first
excited state $|1\rangle$. The Coulomb effects, the ground state
depletion, and the contribution of higher bound states those are
omitted in the SFA, have been accurately incorporated into the
expression of $D(t)$.

%%%%%%%%%%%%%%%%%%%%%%%%%%%%%%%%%%%%%%%%%%%%%%%%%%%%%%%%%%%%%%%%%%%%%%%%%%%%%%%%%%%%%%%%%
\begin{figure}[t]
\begin{center}
\rotatebox{0}{\resizebox *{8.0cm}{8.0cm} {\includegraphics
{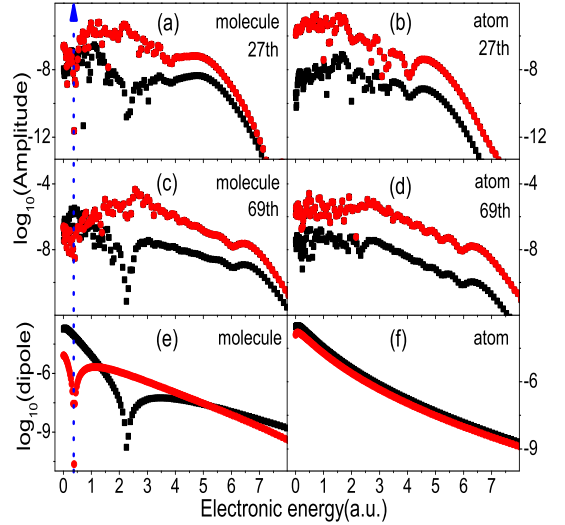}}}
\end{center}
\caption{The contribution of $|G_{n'}(\mathbf{p},\omega)|^2$ to
certain single harmonic orders [(a)-(d)]  and the related transition
dipoles $\langle n'|\nabla V|\mathbf{p}\rangle$ [(e)-(f)] for the
spectra in Fig. 1(a) and (b).  The red curves denote the results
obtained with $n'=0$, the black curves denote those obtained with
$n'=1$. } \label{Fig. 2}
\end{figure}
%%%%%%%%%%%%%%%%%%%%%%%%%%%%%%%%%%%%%%%%%%%%%%%%%%%%%%%%%%%%%%%%%%%%%%%%%%%%%%%%%%%%%%%%%%

The Fourier transformation of $D(t)$ gives %the HHG generated from the dipole moment $D(t)$,
the complex amplitude describing the HHG at the frequency $\omega$
\begin{eqnarray}
F(\omega)=\int dt D(t)e^{i\omega t}=\int
d\mathbf{p}G(\mathbf{p},\omega),
\end{eqnarray}
where $G(\mathbf{p},\omega)=\sum_{n'}G_{n'}(\mathbf{p},\omega)$
and
\begin{eqnarray}
G_{n'}(\mathbf{p},\omega)=\langle n'|\nabla V|\mathbf{p}\rangle\int
dt a^*_{n'}(t)e^{i\omega t}c_\mathbf{p}(t).
\end{eqnarray}
Here, $\langle n'|\nabla V|\mathbf{p}\rangle$ is the transition
dipole matrix element between the continuum $|\mathbf{p}\rangle$
and the bound state $|n'\rangle$.

Eq. 2 shows  that the contributions to  one harmonic $\omega$ come
from the "returned" continuum electrons with varied energy
$E_\mathbf{p}$, each of which is weighted by the probability
amplitude $G(\mathbf{p},\omega)$. The HHG picture revealed here is
somewhat different from the semiclassical ones. The electronic
momentum in the semiclassical
models is the instant momentum that is %subjected to
modulated by the laser field. The momentum in Eq. 2 is the
"laser-field-free" momentum. Thus, Eq. 2 implies a "static" quantum
transition picture of the HHG, i.e., the transition of the continuum
electron $|\mathbf{p}\rangle$ with the time-integrated amplitude
$G_{n'}(\mathbf{p},\omega)$ back to the bound state $|n'\rangle$.
From the "static" picture, we can read some important messages of
the recombination, as  shown below.

In Fig. 1(a) and (b), we compare the harmonic spectra of  1D H$_2^+$
and its reference atom, calculated using  Eq. 1 and Eq. 2,
respectively. We expect that the comparison can give an insight into
the mechanism of the HHG\cite{cyj}. First, we see the black curve in
Fig. 1(a) for H$_2^+$ shows a pronounced suppressed region, i.e., a
broad hollow around the 27th order  indicated by a dotted arrow.
This striking hollow structure is a two-center interference
characteristic of the molecular HHG spectrum and will be discussed
in detail below.
In addition, the red curves in Fig. 1(a) and (b) %calculated from Eq. 2
basically reproduce the behavior of the corresponding black curves
in the plateau region. This verifies the applicability of %our simulation of
Eq. 2. The difference  in  the low energy region is due to the
omission in %our formulation of $D(t)$,
Eq. 2, the bound-bound transitions. Especially, in Fig. 1(a), the
green curve obtained with $n'=0$ (i.e., ignore the first excited
state) shows a deeper hollow than the red one  that includes the
first excited state. In Fig. 1(b) of the atomic case, the two curves
are also analogous. Thus, Fig. 1(a) and (b) reveal %one of our main results
a main result of this present paper that the first excited state,
which  is  usually omitted in  the HHG, could play an important role
in the molecular HHG in the hollow regime.

Next,  we explore the quantum mechanism behind these phenomena. Fig.
2 plots the contribution of $|G_{n'}(\mathbf{p},\omega)|^2$ to
certain single harmonics and the corresponding transition dipoles
$\langle n'|\nabla V|\mathbf{p}\rangle$ for the spectra in Fig. 1(a)
and (b). We choose two typical harmonic orders, the 27th and 69th,
that locate at the interference regime and plateau regime of the HHG
spectrum in Fig. 1(a), respectively.  Fig. 2(a)-(d) obviously show
that for the atomic cases, the amplitude is mainly contributed from
the electrons that return to the ground state. For the molecular
cases, electrons that transit back to either the ground state or the
first excited state could contribute the amplitude significantly,
especially in the low energy regime.

The red curves of $|G_{0}(\mathbf{p},\omega)|^2$  in Fig. 2(a) and
(c) of the molecular cases  show a dip structure around energy
$E(\mathbf{p})=0.35$ a.u., indicated by a blue arrow. The black
curves of $|G_{1}(\mathbf{p},\omega)|^2$ show similar structures
except that the position of the dip structure shifts to a higher
electronic energy of $2.25$ a.u.. Because its amplitude is small, we
can safely ignore its contribution in the following discussion. The
first dip that locates at $E(\mathbf{p})=0.35$ a.u. is important and
corresponds to the pole of the molecular transition dipole in Fig.
2(e). The latter is believed to be relevant to two-center
interference of the diatomic molecules\cite{cyj,Muth-Bohm}.  As we
integrate the amplitude over electron energy, this "two-center
interference effect" remains visible for the case of the 27th,
whereas it vanishes for the case of the 69th, as presented in Fig.
3(a)-(d).

%%%%%%%%%%%%%%%%%%%%%%%%%%%%%%%%%%%%%%%%%%%%%%%%%%%%%%%%%%%%%%%%%%%%%%%%%%%%%%%%%%%%%%%%%
\begin{figure}[t]
\begin{center}
\rotatebox{0}{\resizebox *{8.5cm}{8.0cm} {\includegraphics
{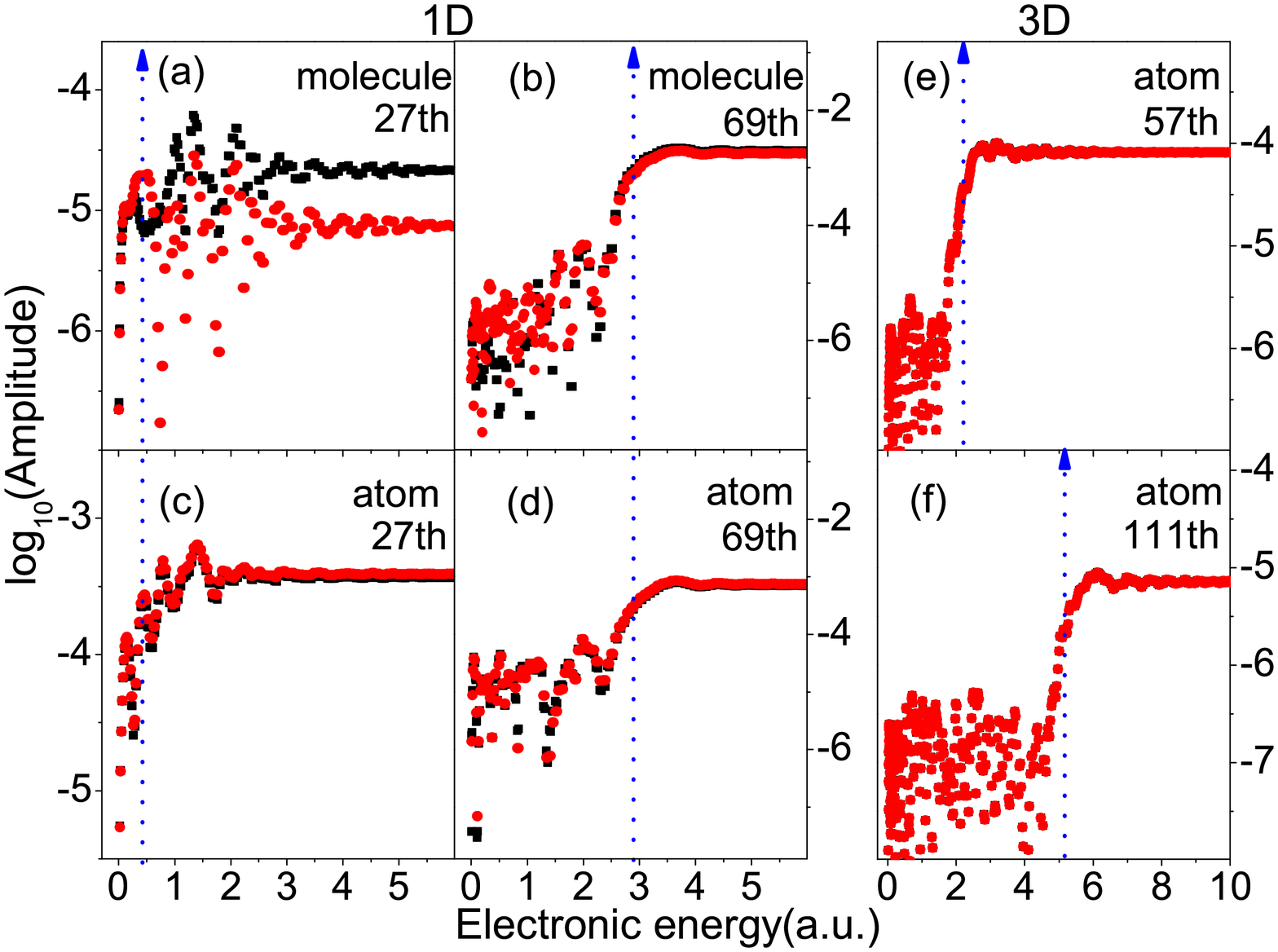}}}
\end{center}
\caption{The contribution of $|\int_{0}^{\mathbf{p}}
d\mathbf{p}G(\mathbf{p},\omega)|^2$ to certain  single harmonic
orders for the spectra  in Fig. 1(a),(b) and (d). The red curves
denote the results obtained with $n'=0$, the black curves denote
those obtained with $n'=0,1$.} \label{fig.3}
\end{figure}
%%%%%%%%%%%%%%%%%%%%%%%%%%%%%%%%%%%%%%%%%%%%%%%%%%%%%%%%%%%%%%%%%%%%%%%%%%%%%%%%%%%%%%%%%%

In the left panels of Fig. 3, we plot the integrated contribution of
$|\sum_{n'}\int_{0}^{\mathbf{p}}
d\mathbf{p}G_{n'}(\mathbf{p},\omega)|^2$. For the 27th order, in
Fig. 3(c) of the atomic case, the red curve  shows that the primary
contributions of $|\int_{0}^{\mathbf{p}}
d\mathbf{p}G_{0}(\mathbf{p},\omega)|^2$ to this order come from
those electrons with a broad energy distribution centered at
$E_\mathbf{p}$=$0.467$ a.u., as indicated by a blue arrow. This
energy value agrees with the energy relation
$\omega=27\omega_{0}=E_\mathbf{p}+I_p$ ($\omega_{0}$ is the laser
frequency)\cite{ilzn04,Levesque}. In addition, the black curve of
$n'=0,1$ almost superposes the red one. In Fig. 3(a) of the
molecular case, the red curve  says that the primary contributions
come from those electrons with energy smaller than
$E_\mathbf{p}$=$0.467$ a.u.. In Fig. 2(a) there is  a dip in the red
curve of $|G_{0}(\mathbf{p},\omega)|^2$ around $E_\mathbf{p}=0.35$
a.u.. We conclude that  two-center interference responsible for the
dip will induce a strong suppression on  this order. Due to the
suppression effect, the first excited state will also play an
important role in
this order. As we can see, the black curve  is higher than the red curve here and %the yield of
this order in Fig. 1(a) is strongly suppressed.

For the 69th order, both Fig. 3(b) and (d) show that the primary
contributions  come from the electrons with a broad energy
distribution centered at $E_\mathbf{p}=2.92$ a.u., as indicated by a
blue arrow. This energy value also agrees with the  energy relation
$\omega=E_\mathbf{p}+I_p$, but is far from the center of the dip in
the red curve in Fig. 2(c). Then we  can expect that two-center
interference has a trivial influence on this order, and the first
excited state plays a negligible role. In Fig. 3(b) of the molecular
case, the black curve almost superposes the red curve and this order
in Fig. 1(a) is not suppressed.

From  Fig. 3(b)-(d), we see that around the blue arrows, there
exists a wide energy regime where the integrated amplitudes increase
rapidly. For example, for the 69th order, the regime spreads from
$E_\mathbf{p}=2.5$ a.u to $3.5$ a.u., which converts to the energy
of about 17 field-photons.  This  reveals an important result that
continuum electrons with a broad energy distribution contribute
equally to one harmonic. The underlying physics is that, from the
quantum viewpoint, the recombination electrons probably emit and
absorb additional energy from the laser field in the vicinity of the
cores\cite{cyj2}. In particular, the blue arrows here indicate the
energy $E_\mathbf{p}=\omega-I_p$. This energy corresponds  to the
semiclassical energy prediction of $E[\mathbf{k}(t)]=\omega-I_p$,
obtained in the SFA for the instant momentum
$\mathbf{k}(t)$\cite{Lewenstein}. Therefore, our simulations
presented here provide a quantitative comparison between the
full-quantum energy distribution and the SFA. The comparison gives
significant suggestions for proper understandings of the HHG in a
full-quantum picture, especially for the intramolecular interference
effect, as shown in Fig. 3(a). Our further analyses show that the
energy distribution is broader for lower harmonic orders, where the
Coulomb potential has a more important role. We expect that the
Coulomb effects are mainly responsible for the physics behind our
simulations. The differences between the SFA and the TDSE have  been
reported in some references\cite{Zhang,Frolov,tong}. In addition,
since the center of the broad energy distribution locates at
$E_\mathbf{p}=\omega-I_p$, it confirms  the relation
$\omega=E_\mathbf{p}+I_p$, which holds in an averaging process as
shown in Ref.\cite{ilzn04,Levesque}.

From the above discussions, we anticipate that the two-center
interference effect in the molecular HHG not only relies on the
population of the first excited state, but also could be affected by
laser parameters such as intensity. This is shown in Fig. 1(c).
First, the plottings in Fig. 1(c) verify the important role of the
first excited state in the molecular HHG. Furthermore, the
comparison between the black curves in Fig. 1(a) and (c) reveals
that the pronounced minimum in the striking hollow region shifts as
the laser intensity changes. Specifically, it shifts from the 27th
order in Fig. 1(a) to the 23rd order in Fig. 1(c), as indicated by
the vertical arrows. It should be mentioned that in
Ref.\cite{lein1}, a spectrum-smoothing procedure is used to help the
identification of the interference minima. This procedure is not
adopted in our analysis, since we %think it somehow ambiguous and we
expect a "direct" comparison of the numerical observations and the
experimental measurements.

The above interference picture is somewhat different from the
point-emitter picture. If we consider that only the electrons, that
transit back to the ground state and have the energy % of %$E_\mathbf{p}$ agreeing with
 $E_\mathbf{p}=\omega-I_p$, contribute significantly to one
harmonic $\omega$\cite{lein1}, Eq. 2 can be approximated to
$F(\omega)=\langle 0|\nabla V|\mathbf{p}\rangle\int dt
a^*_{0}(t)e^{i\omega t}c_\mathbf{p}(t),$ with
$\omega=E_\mathbf{p}+I_p$. The above expression explicitly shows
that the interference minimum in the molecular HHG spectrum that
corresponds to the minimal extremum of the transition dipole
$\langle 0|\nabla V|\mathbf{p}\rangle$ is independent of the laser
intensity.

In the full expression of Eq. 2, it says that the electrons, that
transit back to varied bound states and have diverse energy,
contribute to one harmonic. Our $ab\ initio$ calculations reveal
that the primary contributions to one harmonic come from the
recollision electrons with a broad energy distribution, each of
which is weighted by the probability amplitude
$G(\mathbf{p},\omega)=\sum_{n'}\langle n'|\nabla
V|\mathbf{p}\rangle\int dt a^*_{n'}(t)e^{i\omega t}c_\mathbf{p}(t)$.
This amplitude depends on the laser intensity through the terms
$a_{n'}(t)$ and $c_\mathbf{p}(t)$. The first excited state also has
a nonnegligible role in the molecular HHG. Therefore, the
interference minimum in the molecular HHG spectrum could be
regulated by the laser intensity. We noted that the cutoff in Fig. 2
of Ref.\cite{vozzi} is probably at the 45th order that corresponds
to the laser intensity of $I=3\times 10^{14}$ W/cm$^2$. Compared to
the  intensity of $I=2\times 10^{14}$ W/cm$^2$ used in
Ref.\cite{Tsuneto}, we expect that the uncertain in the calibration
of the laser intensity can occur there and the different
observations in Ref.\cite{Tsuneto,vozzi} can be attributed to the
dependence of the interference pattern on the laser intensity.

We now extend our considerations to the 2D and 3D cases. For the 2D
H$_2^+$, diagonalizing $H_0$ becomes computationally intractable,
and instead of Eq. 2, we simulate using
$F(\omega)=\int\sum_{n'}a_{n'}^*(t)\langle
n'|\vec{\mathbf{e}}\cdot\nabla V|\psi(t)\rangle e^{i\omega t}dt$.
For the 3D hydrogen-like atom, we simulate using $F(\omega)=\int
d\mathbf{p}G(\mathbf{p},\omega)$ with
$G(\mathbf{p},\omega)=\sum_{n',l}^{m=0}G_{n',l,m}(\mathbf{p},\omega)$
and $G_{n',l,m}(\mathbf{p},\omega)=\langle
n',l,m|\vec{\mathbf{e}}\cdot\nabla V|\mathbf{p}\rangle\int dt
a^{*}_{n',l,m}(t)e^{i\omega t}c_\mathbf{p}(t)$. Here,
$|\mathbf{p}\rangle$ is the accurate continuum state with outgoing
wave boundary conditions\cite{tong,Toru}. The definition of $n'$ is
the same as in the 1D case. These results are presented in Fig.
1(d)-(f). For the 3D atom, Fig. 1(d) shows the applicability of our
numerical scheme. Analysis on the integrated contribution of
$|\sum_{n',l}^{m=0}\int_{0}^{\mathbf{p}}
d\mathbf{p}G_{n',l,m}(\mathbf{p},\omega)|^2$ also obtains the
similar results as in the 1D case. We plot the results in the right
panels of Fig. 3. For the  2D molecule, Fig. 1(e) and (f) confirm
the important role of the excited state in the hollow region (as
indicated by the dotted lines) and the shift of the pronounced
minimum (from the 39th to the 51st orders as indicated by the solid
lines) as the laser intensity changes. These major results are also
validated by  our further simulations for H$_2^+$ with varied
 internuclear distances.
Note that this shift arises from not only the nontrivial role of the
excited state in the molecular HHG but also the fact that continuum
electrons in a broad energy regime contribute importantly to one
harmonic. Even we only consider the transition back to the ground
state, the corresponding green curves in the molecular cases in Fig.
1 also reveal the shift of the pronounced minimum. Particularly, all
of our 1D and 2D results  show that the first excited state already
leaves its unambiguous footprints in the molecular HHG spectrum.
This finding throws a new light in the difference of the spectral
amplitudes for the molecule and its reference atom %, that occures
in the orbital tomography procedure\cite{cyj}.

In summary, we have studied the HHG with a full-quantum treatment.
Our simulations revealed a wide  regime of continuum electronic
energy that contributes significantly to one harmonic, and a
nonnegligible contribution of the first excited state to the
molecular HHG. Consequently, the two-center interference pattern is
influenced by the laser intensity, as consistent with experiments.
These results have important implications for ultrafast imaging of
transient molecules.

This work was supported by  NNSF(No.10725521), 973 research program
No.2006CB921400, 2007CB814800, and grants from the Hong Kong
Research Grants Council and Hong Kong Baptist University.

\end{document}